\documentclass{raa_twocolumn}            

\usepackage{graphicx,times}             
\usepackage[square,numbers]{natbib}
\usepackage{amssymb,amsmath}
\bibpunct{(}{)}{;}{a}{}{,}
\usepackage[pagebackref=true]{hyperref}

\begin{document}

\title{SVOM Science User Support at FSC}

   \volnopage{Vol.0 (202x) No.0, 000--000}      
   \setcounter{page}{1}          

   \author{A. Claret 
      \inst{1,4,*}\footnotetext{$*$Corresponding Authors, these authors contributed equally to this work.}
   \and D. Turpin
      \inst{1}
   \and C. Moreau
      \inst{2}      
   \and J.-C. Thome
      \inst{2}
   \and T. Sadibekova
      \inst{1}
   \and F. Daigne
    \inst{3}
   \and B. Cordier
      \inst{4}
   \and F. Agneray
      \inst{2}
   \and M. Brunet
      \inst{5}
   \and N. Dagoneau
      \inst{4} 
   \and A. Formica
      \inst{4} 
   \and O. Godet
    \inst{5}
   \and D. G\"otz
      \inst{1} 
   \and H. Louvin
      \inst{4}
   \and J. Palmerio
      \inst{1}
   \and F. Piron
      \inst{6} 
   \and J. Rodriguez
      \inst{1} 
   \and T. Roland
      \inst{4} 
   \and K. Tazhenova
      \inst{1}
   }

   \institute{
   Université Paris-Saclay, Université Paris Cité, CEA, CNRS, AIM, F-91191 Gif sur Yvette, France; {\it arnaud.claret@cea.fr}
   \and
            Aix Marseille Univ, CNRS, CNES, LAM, Marseille, France;
        \and
            Sorbonne Université, CNRS, UMR 7095, Institut d’Astrophysique de Paris, 75014, Paris, France;
        \and
            CEA Paris-Saclay, Institut de Recherche sur les lois Fondamentales de l’Univers, 91191 Gif sur Yvette, France; \\
        \and
            IRAP, Université de Toulouse, CNRS, CNES, Toulouse, France;
        \and
            Laboratoire Univers et Particules de Montpellier, Université Montpellier, CNRS/IN2P3, 34095 Montpellier, France;
\vs\no
   {\small Received 2025 December xx; accepted 2026 January xx}}

\abstract{The SVOM mission, a Sino-French collaboration dedicated to Gamma-Ray Bursts (GRBs) and transient sources, began scientific operations in 2025. This paper describes the ground computing infrastructure and user support tools for SVOM’s three observing programs: the Core Program (CP), the General Program (GP), and Targets of Opportunity Program (ToO), the latter two being open to the broader scientific community, provided they collaborate with a mission Co-I. The mission adopts operational roles inspired by Swift, including on-duty scientists such as Burst Advocates (BAs), who validate GRB triggers and coordinate follow-up observations, and Instrument Scientists (IS), who calibrate and validate data for all programs. Users can access observation schedules, public data products, and support tools via the French and Chinese mission centers. The SVOM portal serves as the primary interface for accessing these resources, including a GRB public table, API, and user documentation. This paper serves as a guide for both newcomers and external researchers interested in SVOM scientific operations, focusing on aspects related to the CP.
}
   \authorrunning{A. Claret, D. Turpin, C. Moreau et al.}  
   \titlerunning{SVOM Science User Support at FSC}  

   \maketitle

\section{Introduction}  
\label{sect:intro}

Over the past 20 years, Gamma-Ray Burst (GRB) missions—such as NASA’s Neil Gehrels Swift\footnote{\url{https://swift.gsfc.nasa.gov/}} Observatory \citep{swift}, the Fermi\footnote{\url{https://fermi.gsfc.nasa.gov/}} Gamma-ray Space Telescope \citep{fermi}, and ESA’s INTEGRAL\footnote{\url{https://www.esa.int/Science_Exploration/Space_Science/Integral}} observatory \citep{integral}—have developed highly efficient operational frameworks to address the specific challenges of time-domain astronomy and GRB science operations.
Following the legacy of these earlier missions, the SVOM\footnote{\url{https://www.svom.eu/}} mission \citep{raa-cordier} has adopted the same concept of Burst Advocates (BAs) for its Core Program (CP) scientific operations dedicated to the study of GRBs. BAs are scientists on duty 24/7; they oversee the dissemination of real-time alerts to the worldwide community and coordinate the follow-up observations carried out by ground-based telescopes. BA activities are carried out within the two scientific centers of the SVOM mission—the Chinese Science Center (CSC) \cite{raa-han} and the French Science Center (FSC) \cite{raa-louvin}. BAs operate alongside other categories of users involved in the General Program (GP) and in Targets of Opportunity Program (ToO). On the Chinese side, the CSC provides software tools for preparing GP proposals and generating ToO requests \cite{raa-han}. These tools include up-to-date instrument response files supplied by the Instrument Centers (ICs), observation footprint and signal-to-noise estimators, as well as information on source visibility, background maps, catalogs, and more. The Mission Center (MC) \cite{raa-liu} validates and coordinates the observation requests submitted by PIs for all SVOM observing programs (GP and ToO). The Chinese Control Center is responsible for sending the telecommands to be executed onboard the satellite.
For data collection, on the French side, the VHF network \cite{raa-cordier-vhf} receives the triggers generated onboard SVOM and forwards them to the FSC, where near-real-time data analysis is carried out locally or remotely by Chinese and French BAs. On the Chinese side, the full satellite telemetry is collected by X-band antennas connected to the MC. It is also worth noting that China provides access to its Beidou satellite network to improve the response time of alerts transmitted through the French VHF network, which offers only partial global coverage.
Data products are generated and reprocessed as the pipelines evolve throughout the mission’s lifetime, and they are made available through the Space Science Data Center (SSDC) in China. Data from the French instruments are also distributed through the FSC.
In terms of data analysis, the CSC processes data from the Chinese instruments—Gamma-Ray Monitor (GRM) and Visible Telescope (VT) onboard SVOM, as well as the Chinese Ground Follow-up Telescope (C-GFT) and the Ground Wide Angle Camera (GWAC)—and provides support to the BAs for all Chinese instruments. On the French side, the FSC processes data from the French instruments ECLAIRs and MXT onboard SVOM, as well as from the ground-based French-Mexican Ground Follow-up Telescope (FM-GFT). Three Instrument Centers—one for each French instrument—are connected to the FSC and host experts responsible for monitoring and calibrating the French instruments, as well as supporting the BAs for these instruments. These experts are designated Instrument Scientists (IS), a concept also inspired by the Swift mission. All data are reviewed and, if necessary, reprocessed by the ISs, who are responsible for producing the most accurate and validated science products available to SVOM users within the framework of the three observing programs (CP, GP, and ToO).

Following a GRB alert sequence onboard, the SVOM instruments generate a large volume of observation data which is processed by the on-ground processing pipelines to generate public scientific outputs. These include Notices and Circulars, which are broadcast through the NASA Gamma-ray Coordinates Network (GCN)\footnote{\url{https://gcn.nasa.gov/}}, as well as products derived from both the VHF near-real-time telemetry and the X-band complete telemetry \cite{raa-louvin}. These products are classified according to their level of processing, ranging from raw data (L0) to preprocessed data (L1), calibrated data (L2), and more advanced scientific products (L3). The SVOM Collaboration is committed to making L3 products publicly available in their most up-to-date versions as soon as they are available in the scientific database. In addition, contextual scientific information, such as redshift, will also be made available in a timely manner. All these data products can be accessed through the SVOM portal\footnote{\url{https://portal.svom.org}}, or directly via the GRB public table\footnote{\url{https://fsc.svom.org/ifsc-tools/grb-public}}.

In the following subsections, we provide details on the various scientific products that can be expected from the SVOM GRB program. Section \ref{sec:FGS_orga} begins by describing the general organization of the ground segment and the roles of the key personnel responsible for carrying out the main operational tasks. Section \ref{sec:Sec3} focuses on the validation of scientific products from the gamma-ray burst observation program, starting with Notices and Circulars intended for the global community. Finally, Section \ref{sec:Sec4} introduces the SVOM portal, which provides external users with direct access to CP data products through the GRB public table, as well as additional user support tools. We conclude with a summary in Section \ref{sec:conclusion}.

\section{The French ground segment}
\label{sec:FGS_orga}

\subsection{Global organization of the ground segment}
\label{sect:orga}
\begin{figure*}[ht!]
   \centering
   \includegraphics[width=0.8\textwidth]{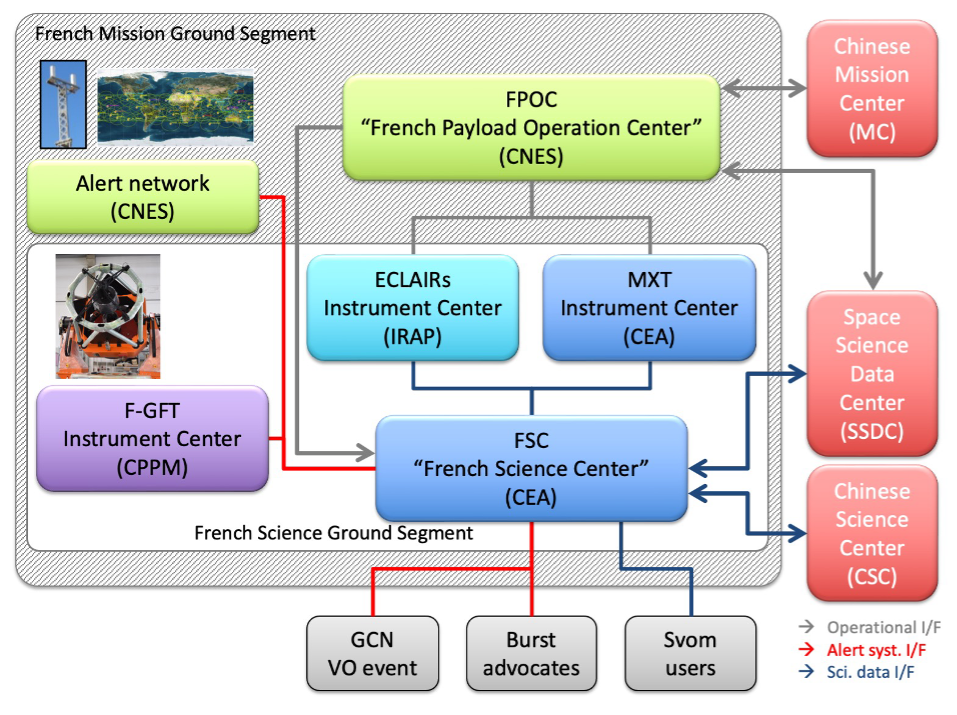}
   \caption{Perimeter of the FSGS. The centers primarily dedicated to operations managed by the Chinese side are shown in red: the SSDC (Space Science Data Center), responsible for collecting data from all instruments via X-band antennas located in China and France, and the MC (Mission Center), responsible for spacecraft operations. On the French side, the FPOC (French Payload Operations Center), shown in green, serves as an interface with both the MC and the SSDC. The FSC is also in direct interface with the Chinese CSC to exchange scientific products. Other acronyms in parentheses are the labs in charge of the corresponding center.}
   \label{fig:FSGS-orga}
   \end{figure*}

In France, most scientific activities are distributed between the French Science Center (FSC) \citep{raa-louvin} and the three French Instrument Centers, each dedicated to a specific French instrument and responsible for monitoring, control, and calibration activities (EIC, MIC, and GIC, for the ECLAIRs, MXT, and FM-GFT instruments, respectively). Those four entities form the so-called French Science Ground Segment (FSGS), shown in Figure~\ref{fig:FSGS-orga}.
The French scientific community at large will interact with the FSGS to follow SVOM mission’s scientific operations.

\begin{figure*}[ht!]
   \centering
   \includegraphics[width=1.05\textwidth]{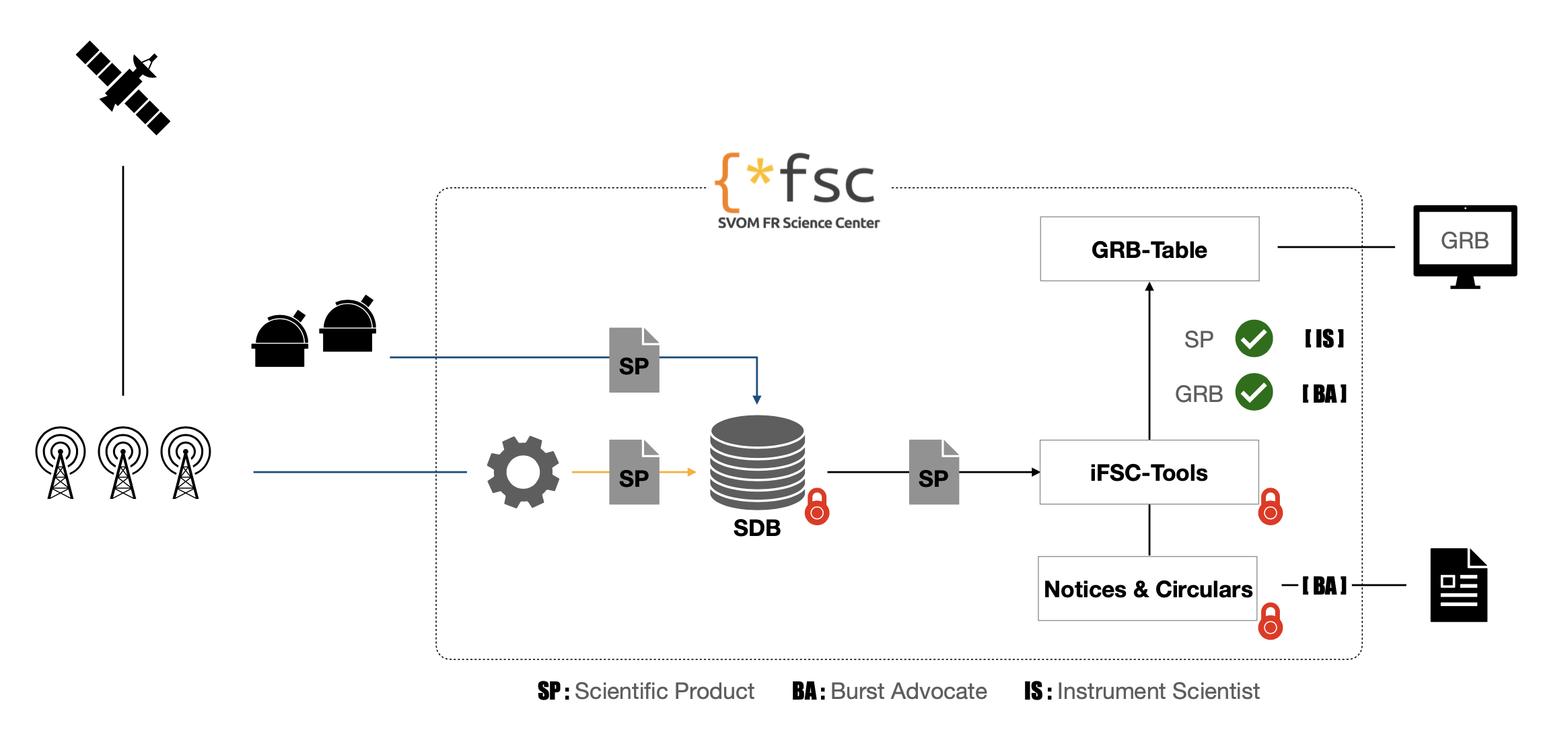}
   \caption{A schematic overview of the data processing, storage, and public distribution of the SVOM Core Program products managed at the French Science Center (FSC). The red padlocks represent data with private access. Further details can be found in \citep{raa-louvin}.}
   \label{FSC_flowchart}
   \end{figure*}

Scientific products are exchanged between the French and Chinese science centers. The ground segment also collects, in near real-time, a subset of data via the alert network provided by CNES, which uses VHF antennas distributed worldwide to cover the satellite’s ground track. On the ground, the alert network is directly connected to the FSC, which is responsible for analyzing this subset of data in real time and distributing the resulting science products. Full data collection via X-band antennas is centralized on the Chinese side, but each science center (FSC and CSC) is responsible for analyzing the complete dataset from its respective instruments. For details on the ground segment architecture and technical implementation, the reader should refer to other papers in this issue \citep{raa-louvin,raa-han}.

\subsection{Scientific operational tasks in SVOM}

In SVOM, different scientific operational roles have been defined to enable the rapid validation of SVOM triggers and the public release of their associated VHF/X-band products with minimal latency. Below, we describe these three main roles, which are slightly adapted from the BA and IS roles originally introduced by Swift:
\begin{itemize}
    \item The \textbf{Trigger BA (BA-T)} is responsible for validating ECLAIRs triggers in real time using the VHF partial dataset, and, when necessary, the full X-band dataset for the more complex cases. The BA-T is also responsible for releasing the \textit{eclairs-wakeup} GCN Notices (see Section \ref{sec:notice}), which are designed to automatically trigger follow-up observations by other instruments. Shortly thereafter, the BA-T takes over the task of writing and sending the high-energy GCN Circular (see Section \ref{sec:Circular}), primarily intended to confirm the GRB detection to the observer community. 
    \item The \textbf{Follow-up BA (BA-F)} is responsible for publicly coordinating SVOM’s multi-wavelength follow-up observations, updating the trigger status in the SVOM GRB public table, and more generally ensuring that the scientific products are made available. If the MXT notice was previously put on hold, the BA-F is also responsible for releasing it.  
    \item The IS are experts in the analysis of instrument data (ECLAIRs, GRM, MXT, and VT). They support the BAs by validating and updating the scientific products derived from the VHF and X-band telemetry. They are also responsible for preparing and distributing both quick-look and refined analyses in the GCN Circulars. For the GRM instrument, which is also capable of triggering GRBs, the GRM-IS is responsible for validating the \textit{grm-trigger}, in collaboration with the ECLGRM-IS (a skilled instrument scientist specialized in ECLAIRs and GRM) during French working hours.
\end{itemize}

As soon as a GRB alert is transmitted to the FSC, both the Chinese and French-Mexican GFTs point toward the potential GRB candidates detected by SVOM to produce results within 5 minutes. In addition, the GWAC collects data from approximately 5 minutes before to 15 minutes after the trigger time. As with all instruments in the SVOM collaboration, dedicated IS for the GFTs and GWAC are responsible for providing the ground-based observation results to the BA-F, who oversees the GRB follow-up.
The alerts issued by the BAs can trigger follow-up observations or counterpart searches by other observatories, both in space and on the ground, including in the multi-messenger domain. The BA is responsible for monitoring all results obtained and disseminated publicly. SVOM BAs also have the authority to request that the Mission Center (MC) in China repoint the SVOM satellite to revisit the most interesting GRBs.

\subsection{Access to data products}

Core Program public scientific products will be made available, accessible through the SVOM portal\footnote{\url{https://portal.svom.org}} hosted in France (see Section \ref{sec:Sec4}). 
The Co-Is of SVOM have access to all scientific products, from raw data files to high-level, model-dependent processed data and results.
In addition, GP and ToO proposals and data access are handled via the Science User Support Services (SUSS) developed at the SVOM CSC. For more details about the CSC and SUSS, see \citep{raa-han}.

\section{Validation of SVOM Core Program data}
\label{sec:Sec3}
Throughout the observation period, satellite data are downlinked via the VHF antenna network, then analyzed and processed in real time by the various instrument pipelines before being stored in the Scientific Database (SDB) at the French Science Center (FSC) \citep{raa-louvin}. Each Gamma-Ray Burst alert is reviewed and validated by the Burst Advocates (BAs) using the burst management and monitoring application (iFSC-Tools). The scientific products (SPs) are then validated by the Instrument Scientists (IS) for each instrument before being released in the public GRB Table.
The resulting scientific products are then made available to the scientific community through the NASA/GCN system for alert dissemination and via the SVOM GRB public table. A quick overview of the SVOM Core Program data-processing flowchart is shown in Figure \ref{FSC_flowchart}. In the following sections, we provide further details on the SVOM public alert system, including the use of NASA GCN Notices and Circulars, as well as the internal procedure for promoting an internal SVOM trigger and its associated products to a public Gamma-Ray Burst event.

   \begin{figure*}[ht!]
   \centering
   \includegraphics[width=1.1\textwidth]{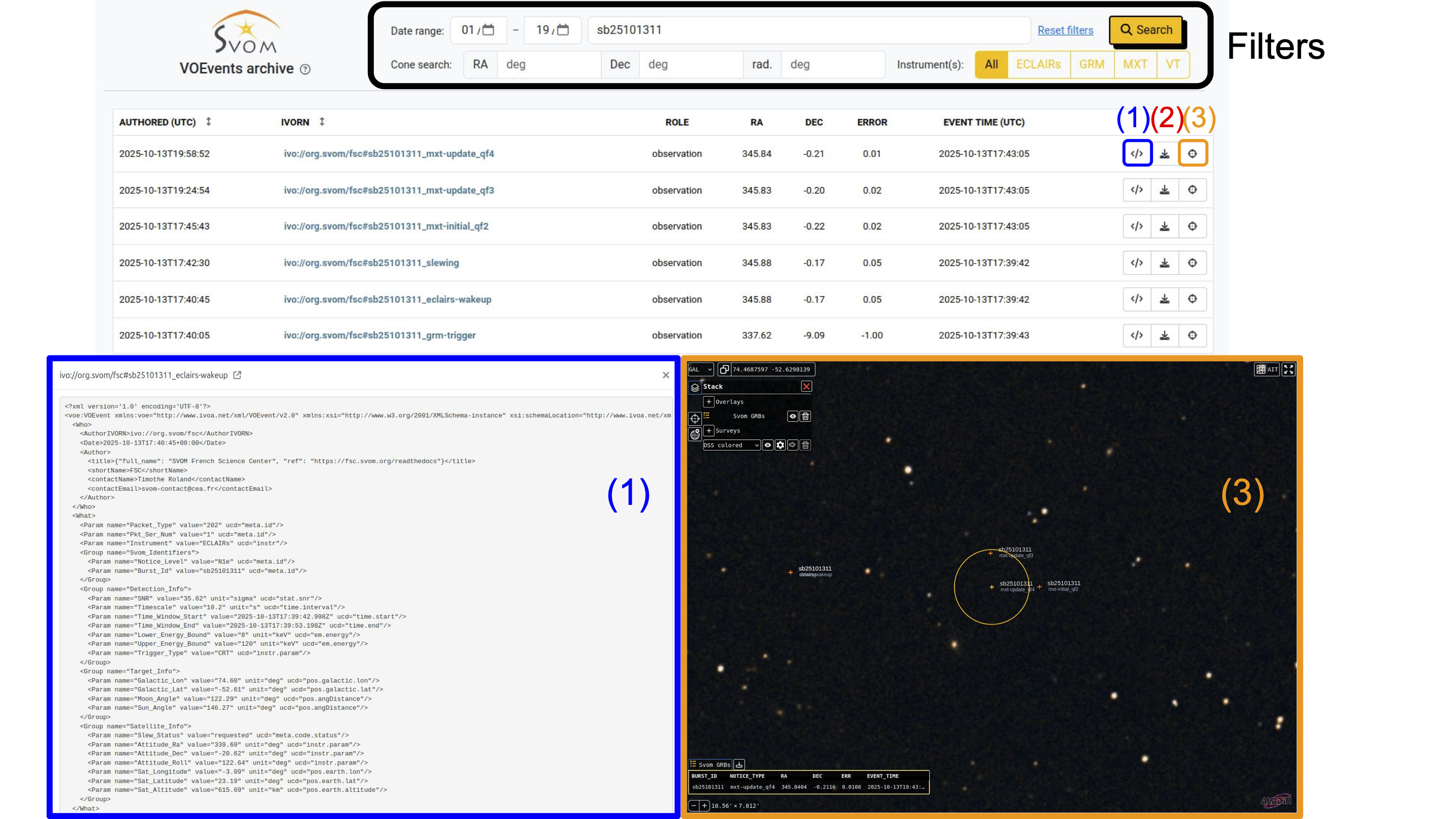}
   \caption{The SVOM public alert page. The filter panel outlined in black at the top highlights the set of filters used to quickly select notices of interest. In this example, we selected all notices generated for the sb25101311 GRB event. The XML content of each notice can be previewed by clicking the button outlined in blue (1) and downloaded using button (2). A customized Aladin Lite view of all selected notices is displayed by clicking the button outlined in orange (3).}
   \label{fig:svom-notice}
   \end{figure*}
   \begin{figure*}[ht!]
   \centering
   \includegraphics[width=1.0\textwidth]{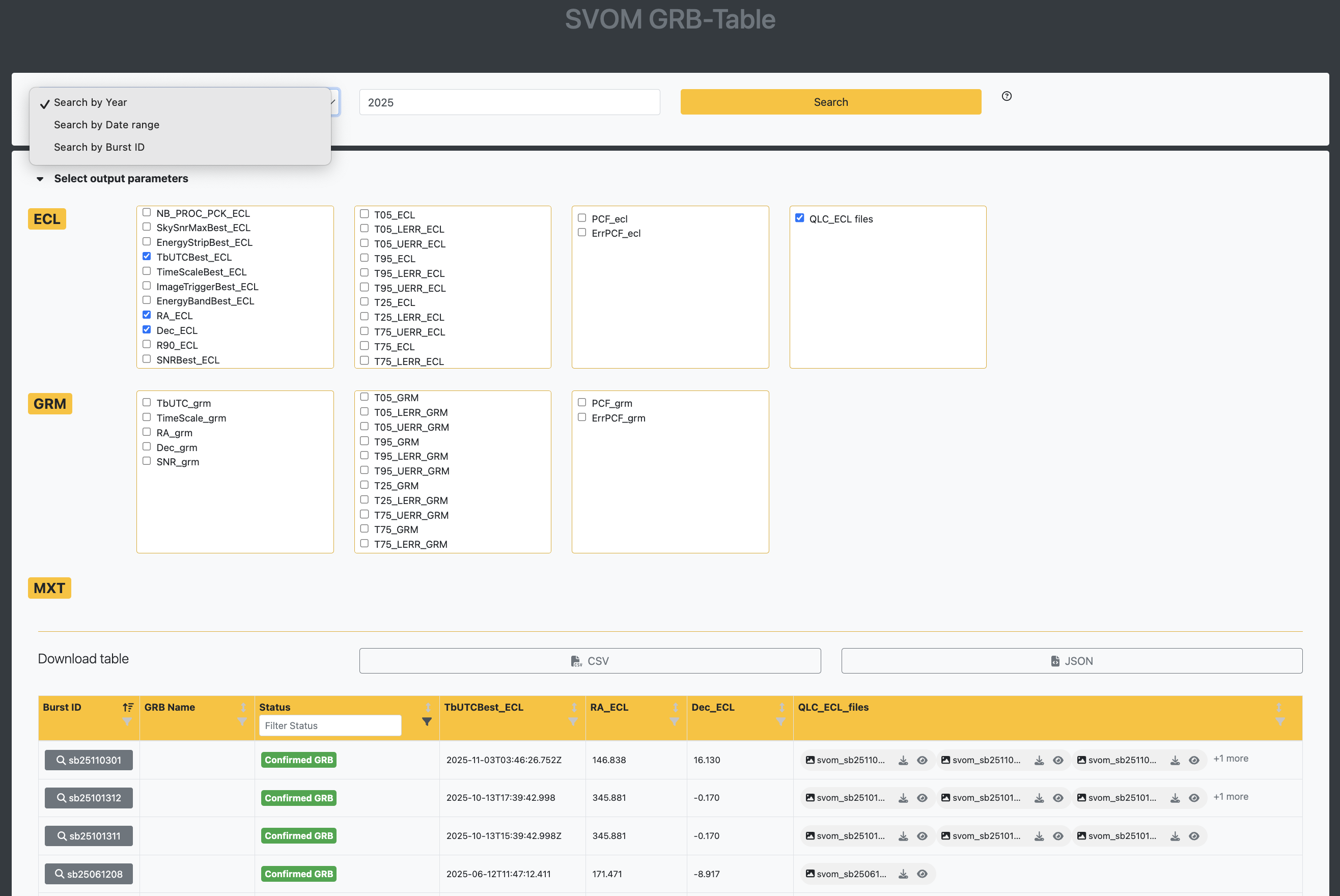}
   \caption{The selection criteria page of the GRB public table allows users to specify which types of products they want to view and download for a customized subset of GRBs. Users are free to tailor their selection according to their needs.}
   \label{GRB-table}
   \end{figure*}

\subsection{Notices}
\label{sec:notice}

The SVOM alerts (from ECLAIRs and GRM), as well as follow-up counterpart positions identified by MXT and VT, are publicly distributed through the NASA/GCN Notices system. For ECLAIRs alerts, the notice types are separated into two categories depending on whether the detected source is spatially consistent with an object in the onboard catalog (\textit{eclairs-catalog}) or not (\textit{eclairs-wakeup}). The list of public notices and Kafka topics is available on the GCN website\footnote{\url{https://gcn.nasa.gov/missions/svom}}. 
More detailed documentation—particularly regarding the notice content and their scientific purpose—is available on the SVOM FSC documentation webpage\footnote{\url{https://fsc.svom.org/readthedocs/svom/notices_and_circulars/notice_levels.html}}. The median latency for the automatic delivery of ECLAIRs and GRM alerts is approximately 13 seconds after the VHF data are received at the FSC \citep{raa-louvin}.

However, the \textit{eclairs-wakeup} notices, as well as the subsequent sequence of notices—\textit{slewing/not-slewing} and \textit{mxt-initial(update)-qf}—may occasionally be issued with longer delays when human validation by the BA-T is required. This can occur in cases of low-SNR detections or when the data are acquired under unusual conditions, such as the satellite passing through regions with high charged-particle backgrounds or pointing towards bright known X-ray sources in the Galactic plane. In contrast to the very short latencies achieved for automatic broadcasting, human-validated \textit{eclairs-wakeup} notices are generally distributed within $\lesssim$ 10 minutes after the VHF data have been received at the FSC.

For the \textit{grm-trigger} notices, automatic distribution occurs only when the instruments are operating outside the SAA regions, where the high charged-particle background can significantly increase the false-alarm rate. This procedure for both ECLAIRs and GRM alerts was established to prevent the dissemination of excessive false alerts to the scientific community. 

The SVOM collaboration provides a public page hosted at the FSC for visualizing the notice sequences produced for ECLAIRs and GRM triggers (see \url{https://fsc.svom.org/alerts/}). A set of filters is available to quickly retrieve the notice sequence for a given burst, or to extract groups of notices issued either for sources located in a selected sky region or within a specified time interval. An example of such a page is shown in Figure \ref{fig:svom-notice}, where we display all notices associated with the event having Burst\_Id\footnote{Internal (SVOM) burst identifier} sb25101311.

\subsection{Circulars}
\label{sec:Circular}

For human readers, we also provide GCN Circulars\footnote{\url{https://gcn.nasa.gov/circulars}}
 with varying latencies, depending on the scientific requirements:
 \begin{enumerate}
    \item A \textbf{High-energy} Circular (\textbf{$\mathbf{T-T_{alert} \leq 30}$ min}): Issued by the BA-T and the ECL/GRM IS to report ECLAIRs detections of confirmed GRBs or X-ray transients, including information on GRM and MXT (non-)detections.
    \item The \textbf{Quick ECLAIRs/GRM light curves} Circulars (\textbf{$\mathbf{T-T_{alert} \lesssim 2}$ hours}): produced by the ECL/GRM-IS to report a rapid analysis of the ECLAIRs and GRM VHF light curves.
    \item A \textbf{Retraction} Circular (\textbf{$\mathbf{T-T_{alert} \leq 2}$ hours, as a goal}): Produced by the BA-T to publicly announce the retraction of an event that had previously generated an automatic eclairs-wakeup notice or a prior high-energy GCN Circular.
    \item The \textbf{Optical follow-up} Circulars (\textbf{$\mathbf{T-T_{alert} \leq 2 - 12 hours}$}): produced by the VT (using either VHF or X-band data), C-GFT and FM-GFT IS to report optical candidate counterparts or flux upper limits.
    \item The \textbf{Refined} Circulars (\textbf{$\mathbf{T-T_{alert} \sim 1-2}$ days}): produced by the IS to report a refined analysis of the temporal and spectral parameters of the ECLAIRs and GRM events, as well as the MXT transient candidates.
    \item The \textbf{GRM} Circulars (\textbf{$\mathbf{T-T_{alert} \sim 1-2}$ days}): produced by the GRM-IS to report the detection of an event that triggered the GRM instrument.
    \item The \textbf{Offline trigger detection} Circulars (\textbf{$\mathbf{T-T_{alert} \sim 1-3}$ days}): Produced by the IS to report ECLAIRs and/or GRM transient detections identified through offline analysis (see Section \ref{sec:OFTG}), either using the full telemetry data) or by replaying VHF data with looser event selection criteria \citep{raa-schanne}.
\end{enumerate}

\subsection{Astronomical telegrams}
\label{sec:ATeLs}

When an onboard alert is identified as originating from a known (catalogued) source, or when the Burst Advocates (BAs) have ruled out a GRB, the alert is assigned to the General Programme / Observatory Science Working Group and to the serendipitous high-energy source group \citep{raa-coleiro}. Based on the scientific relevance of the event, scientists may decide to notify the community by issuing an Astronomer’s Telegram (ATel\footnote{\url{https://astronomerstelegram.org/}}). These alerts typically concern the detection of unexpected flux levels or state variability from known catalogue sources, or the discovery of a new, non-GRB high-energy transient. Such events may also be identified through offline, ground-based analysis of the X-band data, and usually require a re-inspection of the data around the time of the alert.
Consequently, the latency is longer than that associated with the release of GCN notices for GRBs, and may extend to one or two days. This involves:
\begin{enumerate}
    \item  \textbf{Known classes of X-ray/$\gamma$-ray transients} such as X-ray binaries, Tidal Disruption Event (TDE) candidates, flaring or variable stars, active AGN and blazars, and similar sources. 
    \item \textbf{X-ray transients from unknown origin} for which a GRB origin has been ruled out.
\end{enumerate}

\subsection{Offline trigger}
\label{sec:OFTG}

The offline trigger pipeline complements the onboard trigger software. It uses different settings and detection methods based on temporal and imaging excesses, taking advantage of full knowledge of the instrument environment and significantly greater computational resources than those available onboard (see Section 3.1.2 in \citep{raa-coleiro}). 
The offline trigger is able to detect a wide variety of X-ray transients, with durations ranging from a few milliseconds (e.g. Terrestrial Gamma-ray Flashes (TGFs), magnetar flares) to hundreds of seconds (e.g. GRBs, stellar flares). 

The offline trigger can be run either on the full telemetry dataset or on selected time intervals. It is highly configurable, allowing optimization for the detection of various classes of multi-messenger transients under investigation. The pipeline relies on a source catalog to subtract the contributions of known sources from the images and to search for variability in their emission. A machine-learning–based detection module is currently under development. When a new source is detected, the scientific community is informed via an alert notice..

\subsection{Access to SVOM Core Program Scientific Products via the GRB Public Table}
\label{sec:CP_data_provision}

After BA-T validation of the ECLAIRs triggers and the publication of the high-energy GCN Circular, the BA-T and BA-F can promote prompt GRB candidates to the status of "Possible GRB" or "Confirmed GRB", typically within an hour of the alert. This action automatically adds a new entry for the event in the public GRB table (see Section \ref{sec:CP_public_table}).

The validation of scientific products—and their subsequent inclusion in the GRB Public Table—may involve additional delays. Before the full telemetry is received, typically a few hours after the VHF data downlink, the IS validate the VHF-based products and update them as needed using more in-depth analyses. Within 24 hours of a trigger being confirmed or classified as a probable GRB, the VHF-based products are made available to the entire scientific community. High-level (L2) products are also automatically processed and stored in the SVOM Scientific Database (SDB) once the X-band telemetry is received \citep{raa-louvin}. Science-ready (L3) products are produced manually by the IS on duty and are added to the GRB Public Table once validated. The first refined X-band analysis results are typically available within one to two days after the event trigger, with further updates occurring over the following days to weeks, depending on the complexity of the analysis.

Finally, additional information about the event, provided either by external groups or through analysis by our IS, may be manually added to the GRB public table by the BA-F. This includes, for example, redshift measurements, host galaxy properties, the presence of a supernova (SN) or kilonova (KN), and the GRB classification (Type I or Type II). These manual entries are made on the fly by the BA-F as soon as the relevant data become available.

   \begin{figure*}[ht!]
   \centering
   \includegraphics[width=1.0\textwidth]{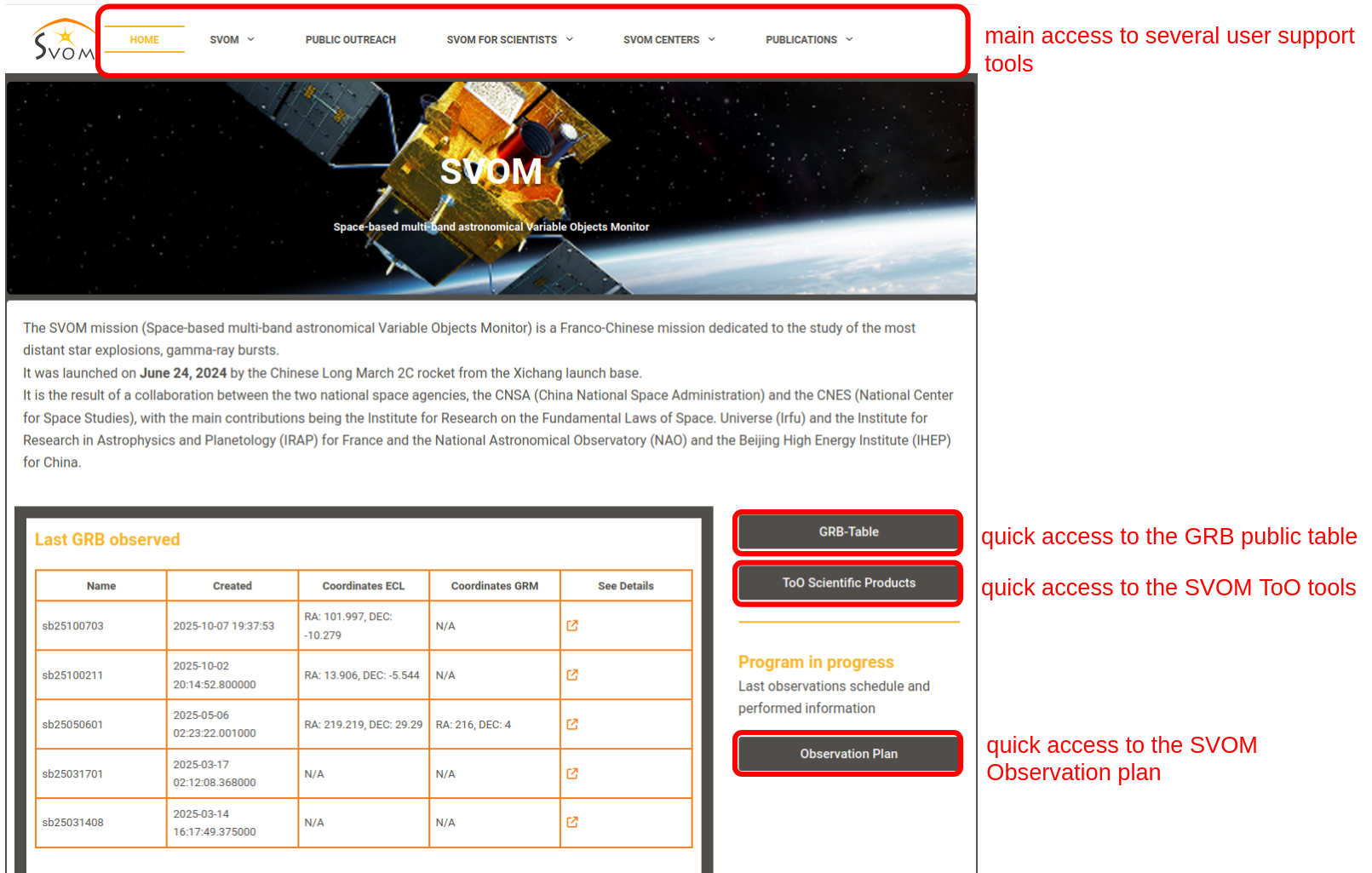}
   \caption{The SVOM Public Portal (current version at the time of writing) is a web interface that provides the scientific community with access to SVOM Core Program public data. Additional links allow external users to navigate to the General and ToO tools. The general public and the press can find detailed information about the mission and the latest news on the \textit{Public Outreach} page. The list of mission publications is also easily accessible via the \textit{Publications} page.}
   \label{fig:svom-portal}
   \end{figure*}
   \begin{figure*}[ht!]
   \centering
   \includegraphics[width=0.99\textwidth]{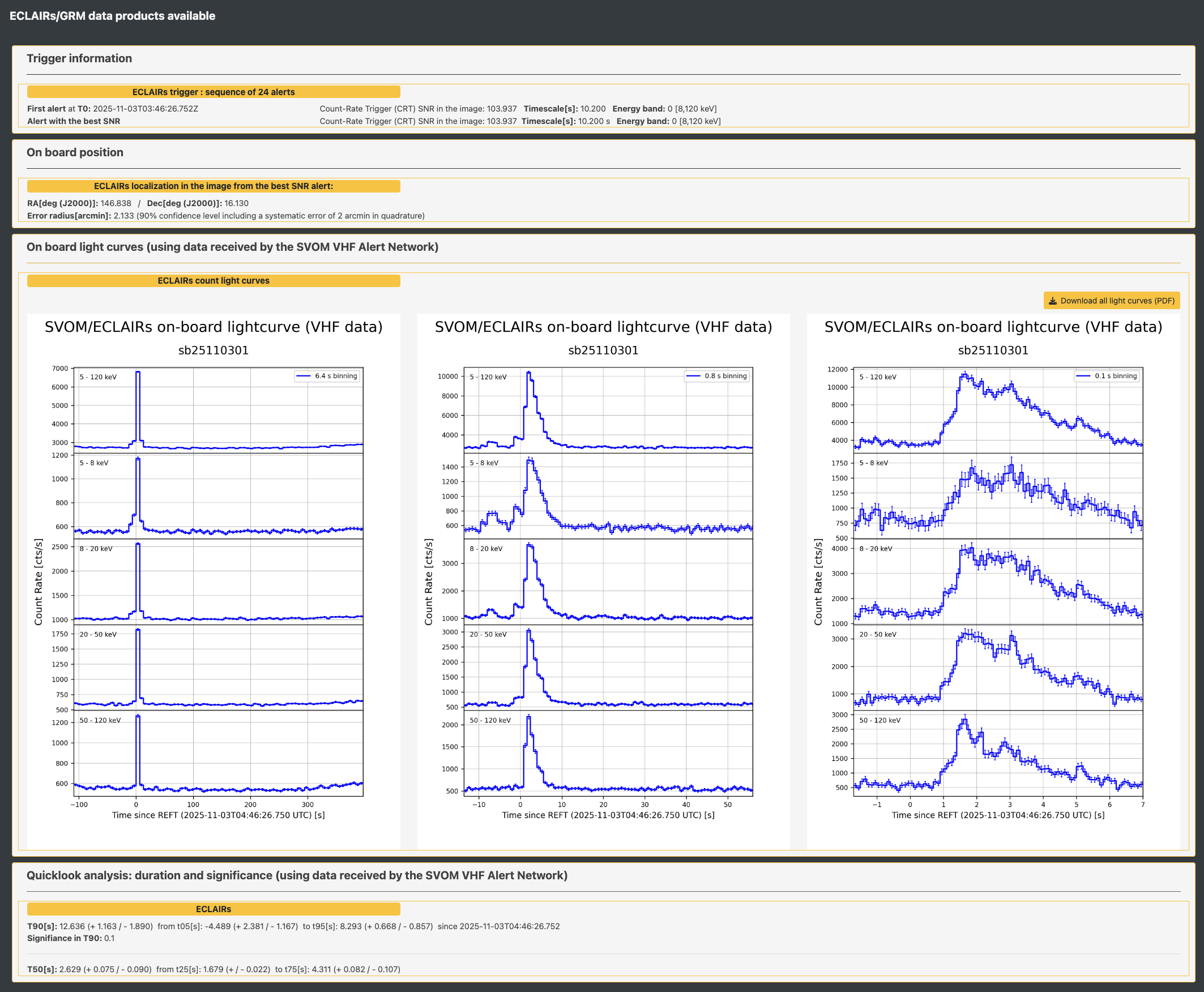}
   \caption{A detailed view of the per-burst information displayed in the GRB Public Table. All of these products and their meta-data can be downloaded in CSV, JSON, FITS, PNG, and PDF formats.}
   \label{GRB-table-detail}
   \end{figure*}

\section{The SVOM Portal}
\label{sec:Sec4}

The SVOM portal is the main web interface made available to the scientific community as well as to the general public (see Figure \ref{fig:svom-portal}). It provides access to a wide range of services, which are briefly described in the following sections.

\subsection{The Gamma-ray Burst public table}
\label{sec:CP_public_table}

The purpose of the \textbf{Gamma-Ray Burst public table} is to provide a web interface that allows non-members of SVOM to access public scientific products, specifically for events classified as possible or confirmed GRBs (see Section\ref{sec:CP_data_provision}).

\subsubsection{Scientific Requirements}

During the alert sequence following a detection on-board SVOM: the GRB public table must provide to the scientific community complementary and first refined information compared to the content of the GCN notices and circulars. Note that only the possible and confirmed GRBs appear in the GRB Public Table with validation procedure described in Section \ref{sec:CP_data_provision} has been completed. The SVOM Circulars should include a link to the GRB Public Table for a quick access. Beyond the alert sequence, the GRB public table must provide to the GRB community a list of all SVOM confirmed GRBs with associated public up-to-date science products. We also provide tools to search for GRBs detected both by ECLAIRs and GRM with specific criteria . This GRB Public Table can be enriched with additional information on SVOM GRBs not directly generated from SVOM data (e.g. redshift).

\subsubsection{Scientific Product Access}

The data are accessible through the GRB Public Table web interface (\url{https://fsc.svom.org/ifsc-tools/grb-public}). Several features are available to help users refine their searches and manage the display of results.

For example, on the main page of the SVOM GRB Public Table, users can select output parameters and export the table.
They can currently download the GRB table in CSV or JSON format, and additional formats (FITS, etc.) are planned.
Users can also search for GRBs using various criteria, such as by year, date range, or directly by Burst\_Id. The range of possible search criteria will expand as new output parameters become available in the GRB table (all features shown in Figure~\ref{GRB-table}).
For each GRB with available light curves, users can download them individually or view them online. They will also be able to download all available products for a given GRB. Scientific products can be downloaded in CSV, FITS, PNG, or PDF format, depending on the product type.

By selecting a GRB in the table, users can display all available scientific products in detail, including trigger information, onboard position, onboard light curves, and quicklook analysis such as duration and significance ( Figure~\ref{GRB-table-detail}).
Questions about a given GRB and its associated scientific products can be sent to the BA in charge, whose email address is provided on the page of the GRB in the GRB Public Table.
Users can also reach the application support team through the contact form available in the GRB Public Table interface.

An API is currently available to retrieve information about the GRBs listed in the GRB Public Table. The results are provided in JSON format.

The required criteria for performing this search are either a start and end date range (e.g., (\url{https://fsc.svom.org/ifsc-tools/server-rest/grb-public/grb-list?start=2025-01-01&end=2025-12-31}) or a Burst\_Id (e.g., \url{https://fsc.svom.org/ifsc-tools/server-rest/grb-public/grb-list?burst_id=sb25111605}). 
As the search options in the web interface evolve, additional criteria may also become available in the API.
Plans are underway to offer, in a future version, the ability to download scientific products directly through the online API.
API usage documentation is available on the GRB-table website.

\subsection{Access to the mission publications}
Through the SVOM Public Portal, users can quickly access scientific publications related to the mission and its instruments, as well as science papers categorized into GRB and Observatory publications.

\subsection{The SVOM website for the press and general public}
The SVOM collaboration has developed a comprehensive website to share news and outreach content related to the mission, its instruments, the scientific objectives, and major discoveries. This website is accessible at the following URL: \url{https://www.svom.eu/en/home/}
, or by clicking the Public Outreach button in the main menu of the SVOM portal (see upper part of Figure \ref{fig:svom-portal}).

\subsection{Links to the SVOM Chinese user support tools}
On the SVOM portal, users can have access to the SVOM workplan, maintained by the Chinese MC. This includes the Pre-Planned Science Timeline (PPST), which outlines the planned observations for SVOM, and the As-Flown Science Timeline (AFST), which provides the log of observations actually performed by SVOM, including uploaded ToOs and GRBs. This tool can be accessed by clicking the Observation Plan button on the SVOM portal (see Figure \ref{fig:svom-portal}) or directly via \url{https://soqt.smoc.ac.cn/}. 

Additional Chinese user support tools, particularly for the General and ToO programs, are available through the SVOM CSC. A link to the CSC/SUSS page (\url{https://www.svom.cn/suss/#/svomPage}
) is also provided on the SVOM portal, as shown in Figure \ref{fig:svom-portal}. For information on user rights and data access via the Chinese user support services, readers are referred to the CSC publications \cite{raa-huang,raa-han}.

\subsection{SVOM User Support Website Map}

The most useful links to know are the following.

\begin{itemize}
\item \textbf{SVOM Portal} \\
The central entry point for the SVOM mission, providing general information and access to mission resources:
https://portal.svom.org/
\item \textbf{French Science Center (FSC) and Documentation} \\
The FSC website offers access to scientific tools, documentation, and data services for SVOM users:
https://fsc.svom.org/
\item \textbf{GRB Public Table} \\
A public interface providing information on detected GRBs and associated products:
https://fsc.svom.org/ifsc-tools/grb-public
\item \textbf{Science User Support Services (SUSS)} \\
The Chinese Science Center portal for SUSS, providing complementary access to SVOM resources:
https://www.svom.cn/suss/\#/svomPage
\item \textbf{SVOM Workplan} \\
The mission operations and workplan interface hosted by the SVOM Mission Center:
https://soqt.smoc.ac.cn/
\item \textbf{SVOM Outreach and Public Website} \\
The official SVOM outreach website for the general public, press, and educational content:
https://www.svom.eu/en/home/
\end{itemize}

\section{Conclusion}
\label{sec:conclusion}

The key aspects of SVOM’s scientific operations have been described in detail, both in terms of organizational structure and the production of scientific products, ranging from Notices and Circulars distributed with minimal latency to the GRB characteristics compiled in the public GRB table. 
This organization quickly proved to be very effective, as demonstrated by the first results on gamma-ray bursts, with reference to the corresponding article in the Special Issue \citep[see][]{raa-daigne}.

The SVOM portal brings together services useful to users, including access to public data from the GRB observation program and the observation schedule for the General Program. From this portal, users can also easily access a separate website that provides detailed information intended for the general public and the press.

Improvements to better meet user needs will be continuously implemented throughout the lifetime of SVOM’s scientific operations. In particular, teams of scientists and developers are working to increase automation and enhance the reliability of scientific products, while minimizing dissemination time for the international community.

\begin{acknowledgements}
The Space-based multi-band astronomical Variable Objects Monitor (SVOM) is a joint Chinese-French mission led by the Chinese National Space Administration (CNSA), the French Space Agency (CNES), and the Chinese Academy of Sciences (CAS). We gratefully acknowledge the unwavering support of NSSC, IAMCAS, XIOPM, NAOC, IHEP, CNES, CEA, and CNRS. We also gratefully acknowledge support from the SVOM FSC engineer team and CNRS/IN2P3 Computing Center (Lyon - France) for providing computing and data-processing resources needed for this work.
\end{acknowledgements}



\begin{thebibliography}{99}


\bibitem[Coleiro et al. (2026)]{raa-coleiro} Coleiro A. et al., \textit{SVOM Observatory Science: results from the first year of the mission}, RAA 2026, 25, this issue

\bibitem[Cordier et al. (2026)]{raa-cordier} Cordier B. et al., \textit{The SVOM mission, its profile and its system}, RAA 2026, 25, this issue

\bibitem[Cordier et al. (2026)]{raa-cordier-vhf} Cordier B. et al., \textit{The VHF alert network of the SVOM mission}, RAA 2026, 25, this issue

\bibitem[Daigne et al. (2026)]{raa-daigne} Daigne F. et al., \textit{First Gamma-Ray Burst Observations with SVOM}, RAA 2026, this issue

\bibitem[Gehrels et al. (2004)]{swift} Gehrels N. et al., \textit{The Swift Gamma-Ray Burst Mission}, 2004, \apj, 611, 1005

\bibitem[Han et al. (2026)]{raa-han} Han X.-H. et al., \textit{SVOM Science User Support Services at CSC}, RAA 2026, this issue

\bibitem[Huang et al. (2026)]{raa-huang} Huang M.-H. et al., \textit{Chinese Science Center Infrastructure}, RAA 2026, this issue

\bibitem[Liu et al. (2026)]{raa-liu} Liu H. et al., \textit{The SVOM Mission Center}, RAA 2026, 25, this issue

\bibitem[Louvin et al. (2026)]{raa-louvin} Louvin H. et al., \textit{The SVOM French Science Center Infrastructure}, RAA 2026, 25, this issue

\bibitem[Mereghetti et al. (2003)]{integral} Mereghetti S. et al., \textit{The INTEGRAL Burst Alert System}, 2003, \aap, 411, L291

\bibitem[Meegan et al. (2009)]{fermi} Meegan C. et al., \textit{The Fermi Gamma-ray Burst Monitor}, 2009, \apj, 702, 791

\bibitem[Schanne et al. (2026)]{raa-schanne} Schanne S. et al., \textit{Overview of the ECLAIRs trigger for SVOM gamma-ray burst detection}, RAA 2026, this issue

\end{thebibliography}

\label{lastpage}

\end{document}